\documentclass[%
preprint,
showpacs,
preprintnumbers,
 amsmath,amssymb,
 aps,prd,
]{revtex4-1}

\usepackage{graphicx}
\usepackage{dcolumn}
\usepackage{bm}
\newcommand{\jcap}{\rm{JCAP}}
\newcommand{\mnras}{\rm{MNRAS}}

\begin{document}

\preprint{DRF/IRFU, Universit\'e Paris-Saclay}

\title{Gravitational lensing of photons coupled to massive particles}

\author{J-F. Glicenstein}
\email{glicens@cea.fr}
\affiliation{%
IRFU, CEA, Universit\'e Paris-Saclay, F91191Gif-sur-Yvette, France \\
}%

\date{\today}

\begin{abstract}
The gravitational deflection of massless and massive particles, both with and without spin, has been extensively studied.
This paper discusses the lensing of a particle which oscillates between two interaction eigenstates. 
The deflection angle, lens equation and time delay between images are derived in a model of photon to hidden-photon oscillations. 
In the case of coherent oscillations, the coupled photon behaves as a massive particle with a mass equal to the product of the coupling constant and hidden photon mass.
The conditions for observing coherent photon-hidden photon lensing are discussed. 
\end{abstract}

\pacs{04.80.Cc - 14.70.Pw - }
\keywords{Gravitational lenses - Particle mixing and oscillation}
\maketitle


\section {\label{sec:Introduction}Introduction} 
The gravitational deflection of  massless and massive particles, with and without spin, has been studied by numerous authors. 
The gravitational deflection of massive photons has been known for several decades \citep{1973PhRvD...8.2349L}.
The deflection angle for a single flavor, massless neutrino was derived by \citet{2005PhRvD..71g3011L} and found to agree with the predictions of General Relativity for massless photons.
The optics and deflection angle for a spin-2 particle has been obtained by \citet{2007PhRvD..75d4022P}.
Several papers have discussed explicitely neutrino lensing \citep{1987Natur.327..375B,2007APh....28..348M,1997PhRvD..56.1895F}.  
In a recent paper, the ANTARES neutrino telescope \cite{2014JCAP...11..017A} has obtained limits on the neutrino emission of 4 lensed quasars.
However, to the best of the author's knowledge, no detailed treatment of neutrino lensing with flavor oscillations, giving the images positions and delays has been published, presumably due to 
the complex formalism describing neutrino propagation in a gravitational field.
The purpose of this paper is to study the lensing of oscillating particles with 
a simpler model where photons are coupled to hidden, massive photon-like particles (HP).   
The most straightforward approach to the lensing of oscillationg particles consists in following indepently the mass eigenstates on their geodesics, then recombine them (coherently or not) at the observer level. This point of view is taken in section   \ref{sec:coherence}.
The other approach is the direct study of the propagation of a photon-HP beam. 
The  mixing of photons to HP is shown in section \ref{sec:opticalindex} to be equivalent to the propagation of photons in a refractive medium. 
An effective refractive index is calculated in section \ref{sec:refindex}.  The Synge formalism \citep{synge1960relativity}, which provides Hamilton's equations for photon propagation in refractive media,
is used in the section \ref{sec:hplensing} to derive the deflection angle and time delay of the mixed photon. Finally, signatures of photon-HP mixing in the deflection of radio waves by the sun are discussed in  section \ref{sec:coherence} .

\section{Effective refractive index of photons mixed with hidden photons}\label{sec:opticalindex}
\subsection{Hidden photon model}
Photons are predicted to couple to massive uncharged particles in several extensions of the standard model of particle physics. The most popular of these particles is the axion, which is a solution to the strong CP problem \citet{1981PhLB..104..199D}. Axion oscillate into photons in external magnetic fields. Axions and other axion-like particles have been extensively searched for and constrained \citep{2008LNP...741...51R,2015JPhCS.632a2004D}.  
The hidden photon (HP)-photon oscillations \citep{2007PhRvD..76k5005A} share many  features with the photon-axion oscillations,but is simpler in several aspects (e.g. the equivalent refractive index in vacuum is isotropic). In this paper, the HP-photon oscillations will be used as a toy model to study the lensing of oscillating particle systems.   
This section uses the notations of \cite{2007PhRvD..76k5005A}.


In the hidden photon model, the photon field $A_{\mu}$ couples with a coupling strength $\chi$ to a massive spin 1 field $\tilde{B}_{\mu}$ with mass $\mu.$  The lagrangian contains the usual electromagnetic lagrangian and a $\tilde{B}_{\mu}$
related part:
\begin{equation}
\label{masssimple}
{\mathcal{L}}_{B}=\frac{1}{2}\mu^2 \left(\tilde{B}^{\mu}\tilde{B}_{\mu}-2\chi \tilde{B}^{\mu}A_{\mu}+
\chi^2 A^{\mu}A_{\mu}\right)
\end{equation}

The equations of motion in the $z$ direction are
\begin{equation}
\label{eq:eom}
\left[(\omega^2+\partial^{2}_{z})\left(
          \begin{array}{cc}
            1 & 0 \\
            0 & 1 \\
          \end{array}
        \right)-\mu^2\left(
               \begin{array}{cc}
                 \chi^2 & -\chi \\
                 -\chi & 1 \\
               \end{array}
             \right)\right]\left(
                             \begin{array}{c}
                               A \\
                               \tilde{B} \\
                             \end{array}
                           \right)=0.
\end{equation}

The dispersion relations of plane waves with energy $\omega$ and momentum $k$ are obtained from equations (\ref{eq:eom})
\begin{equation}
\label{eq:dispersion}
\begin{vmatrix}
\omega^2-k^2 -\mu^2 \chi^2 & \mu^2\chi \\
 \mu^2\chi  &   \omega^2-k^2 -\mu^2
 \end{vmatrix} 
                           =0.
\end{equation}
These dispersion relations have two branches which describe a massless field, the photon, and a massive field with mass $\mu' =\mu\sqrt{\chi^2 + 1}.$

The propagation eigenvectors can be projected in the (A,$\tilde{B}$) interaction basis as
\begin{eqnarray}
\Gamma_1 &= \frac{1}{\sqrt{1+\chi^2}} A +  \frac{\chi}{\sqrt{1+\chi^2}}\tilde B \nonumber\\
\Gamma_2 &= -\frac{\chi}{\sqrt{1+\chi^2}} A +  \frac{1}{\sqrt{1+\chi^2}}\tilde B, 
\end{eqnarray} 
where $\Gamma_1$ is the photon state.

\subsection{Equivalent refractive index}\label{sec:refindex}

In this section, the propagation of a photon beam originating from a cosmic source is studied. Photons and HP are described by plane waves, with energy $\omega,$ and propagating in the $z$ direction. 
In section \ref{sec:coherence}, photon-HP beam coherence effects  will be discussed and it will be more appropriate to use wave packets.
Initially, the photon beam is in a pure $A$ state, which is projected into the propagation basis as:
\begin{equation}
\Phi (t=0)=  A= -\frac{\chi}{\sqrt{1+\chi^2}} {\Gamma}_{2} +  \frac{1}{\sqrt{1+\chi^2}} \Gamma_{1}
\end{equation}
At a later time t, 
\begin{eqnarray}
\Phi (t)=&&  -\frac{\chi}{\sqrt{1+\chi^2}} \Gamma_{2}\exp{i(\omega t -\sqrt{\omega^2-\mu'^2 } z)}\nonumber \\
 && +  \frac{1}{\sqrt{1+\chi^2}}\Gamma_{1} \exp{i\omega(t - z)}. 
\label{eq:Aoft}
\end{eqnarray}
Equation (\ref{eq:Aoft}) assumes that the energy is conserved in the photon-HP oscillations, and momentum not conserved. This procedure is commonly used for neutrino oscillations has been proven to give consistent results  \citep{2009PAN....72.1363A}. 
The photon amplitude A(t) is obtained by projecting $\Phi(t)$ over A. 
 \begin{equation}
A(t) = \Phi(t). A=  \frac{\exp{i\omega(t - z)}}{1+\chi^2} (\chi^2 \exp{i(\omega-\sqrt{\omega^2-\mu'^2 }) z} +  1) 
\end{equation}

Writing $A(t)=\rho  \exp{i\omega(t - z)} \exp{i\phi}$ and assuming $\frac{\mu'}{\omega} \ll 1,$ 
the photon survival probability is
\begin{eqnarray}
\rho^2 &= 1 - \frac{4 \chi^2}{ (1+\chi^2)^2}\sin^2{((\omega-\sqrt{\omega^2-\mu'^2 })z/2)} \nonumber \\
&= 1 - \frac{4 \chi^2}{ (1+\chi^2)^2}\sin^2{(\frac{\mu'^2z}{4\omega})}
\end{eqnarray}
and the phase of the photon beam is
\begin{equation}
\phi 
\simeq \frac{z \chi^2 \mu^2}{2\omega} 
\end{equation}

Defining the effective wave number by $k_{\mathrm{eff}} = \omega - \frac{\chi^2 \mu^2}{2\omega},$
the  phase velocity $v_{\phi}$ and refractive index $n$ are obtained: 
\begin{subequations}
\begin{eqnarray}
v_{\phi} &= \frac{\omega}{k_{\mathrm{eff}}}   
\simeq 1 + \frac{\chi^2 \mu^2}{2\omega^2} \\
\label{eq:nvalue}
n &  = \frac{1}{v_{\phi} } \simeq 1 - \frac{\chi^2 \mu^2}{2\omega^2}  
   \simeq v_{g},
\end{eqnarray}
\end{subequations}
where $v_{g}$ is the group velocity.  The motion of a photon oscillating to an HP is thus similar to the motion of a photon in  a refractive medium or in a plasma.



\section{Gravitational lensing of photons mixed with hidden photons}\label{sec:hplensing}
Gravitational lensing in a refractive medium has been studied by several authors (see e.g \citet{2010MNRAS.404.1790B}) using the formalism of  \citet{synge1960relativity}.
Assume that the light from a distant source propagates in the gravitational field  $U(r) \ll 1$ of a massive lens.
The spacetime is described by the isotropic metric $g_{ij}$, given in cartesian coordinates by:
\begin{subequations}
\begin{eqnarray}
ds^2 &= g_{ij} dx^{i}dx^{j} = g_{00} dt^2 + g_{s s} dr^2 \\
&= -(1+2U(r)) dt^2 + (1-2U(r))dr^2 
\label{eq:1}
\end{eqnarray}
\end{subequations}
In equation (\ref{eq:1}), the speed of light c is set to 1. The metric signature is (-,+,+,+), greek letters are used for the spatial part of the metric
and $dr^2 = dx^{\alpha} dx^{\beta} \delta_{\alpha \beta}$

The Hamiltonian of a photon with energy  $\omega = -p_0 \sqrt{-g^{00}}$  in a medium with refractive index $n(x^{\alpha}, \omega)$ is 
\begin{equation}
H(x^{i}, p_{i}) = \frac{1}{2}(g^{ij}p_{i}p_{j} - (n^2 - 1)\omega^2)
\label{eq:Synge}
\end{equation}
with the further restriction $H = 0.$ The latter is the dispersion relation in the medium of index $n$.  With the metric defined by  (\ref{eq:1}), the dispersion relation can be simplified to
\begin{equation}
\label{eq:dispersionrelation}
g^{ss} \sum_{\alpha} p_{\alpha}^2= - n^2 g^{0 0} p_{0}^2  
\end{equation}

 The Hamilton equations for the $x^{\alpha}$ coordinates of the photon are, in the medium rest-frame
\begin{subequations}
 \begin{eqnarray}
\label{eq:Hamilton1}
\frac{dx^{\alpha}}{d\lambda} &= g^{\alpha\alpha} p_{\alpha} = g^{SS} p_{\alpha} \ \  \ (\mbox{space components)}) \\
\frac{dx^{o}}{d\lambda} &= g^{0 0} p_{0} +((n^2-1)+n \omega \frac{dn}{d\omega}) p_0 g^{00}\nonumber \\
&= (n^2 + n\omega \frac{dn}{d\omega}) p_0 g^{00}, 
\label{eq:Hamilton2}
\end{eqnarray}
where $\lambda$ parametrizes the position of the photon.
\end{subequations}
Introducing the group velocity 
\begin{equation}
\frac{1}{v_g}= \frac{\partial (n\omega)}{\partial \omega}
\end{equation}
and the momentum $P^2 = \sum  p_{\alpha}^2,$ and using the dispersion relation (\ref{eq:dispersionrelation}), equation (\ref{eq:Hamilton2}) can be 
rewritten as
\begin{subequations}
 \begin{eqnarray}
  \frac{dx^{o}}{d\lambda}   & = (n + \omega \frac{dn}{d\omega}) P \sqrt{(g^{SS})\times(-g^{00})} \nonumber \\
                                         & = \frac{P}{v_g} \sqrt{(\sum g^{ss}\times(-g^{00})}
\end{eqnarray}
\end{subequations}

The length element can be calculated using equation (\ref{eq:Hamilton1})
\begin{equation}
\frac{dr}{d\lambda} = g^{SS} P, 
\end{equation}
so that 
\begin{equation}
\frac{dx^{o}}{dr} = \frac{1}{v_g} \sqrt{\frac{-g^{00}} {g^{SS}}} = \frac{1}{v_g} \sqrt{\frac{g_{SS}} {-g_{00}}} 
\end{equation}
The travel time from the source S to the observer O is given by
\begin{equation}
\label{eq:Fermat1}
T  = \int_{O}^{S}{ \frac{1}{v_g} \sqrt{\frac{g_{SS}} {-g_{00}}} dr} \\
\simeq  \int_{O}^{S}{ \frac{(1-2U(r))dr}{v_g}}
 \end{equation}
In equation (\ref{eq:Fermat1}), $v_g$ depends on position through the dependence of the energy of the particle on the gravitational potential $U(r),$ 
For the refractive index obtained in section \ref{sec:refindex},
\begin{subequations}
\begin{eqnarray}
\frac{1}{v_g} &= \frac{\partial (n\omega)}{\partial \omega} = 1 + \frac{\chi^2 \mu^2}{2\omega^2} = 1 - \frac{\chi^2 \mu^2}{2p_0^2 g^{00}} \\
&\simeq (1+\frac{\chi^2 \mu^2}{2p_0^2} +\frac{\chi^2 \mu^2 U(r)}{p_0^2})
\label{eq:potential2}
\end{eqnarray}
\end{subequations}

To first order in $U(r),$ the travel time is given by the integral
\begin{equation}
\label{eq:propagationtime}
T =  \int_{O}^{S} dr (1  +\frac{\chi^2 \mu^2}{2p_0^2} -2 U(r))
 \end{equation}


 A generalization of the Fermat principle  from geometrical optics \citep{synge1960relativity,1992grle.book.....S} is used in the paper to derive the deflection angle. The optical path length from the source to the observer, 
\begin{equation}
\label{eq:Fermat2}
\mbox{\cal{L}} =  \int_{O}^{S}{ n \sqrt{\frac{g_{SS}} {-g_{00}}} dr} \simeq \int_{O}^{S}{ n(1-2U(r)) dr}
 \end{equation}
 is extremal on  light paths.
Taking $n$ from equation (\ref{eq:nvalue}) and developping as in (\ref{eq:potential2}), one obtains to first order in $U(r)$
\begin{equation}
\label{eq:Fermat-th}
\mbox{\cal{L}} =  \int_{O}^{S} dr (1- \frac{\chi^2 \mu^2}{2p_0^2} -2 U(r))
 \end{equation}

In the thin lens approximation, the lensed particle moves 
 on a straight line until it get deflected towards the observer. The angular distance to the lens is $D_{OL},$ the angular distance from lens to observer is 
 $D_{LS}.$ Coordinates are taken in the plane perpendicular to the line of sight (the "lens plane"). The projected source position on the lens plane is at $\eta,$ and the impact parameter of the particle trajectory at $\zeta.$  
  
 The remainder of the paper uses the  Schwarzschild lens model, appropriate for lensing by stars and black holes.  
 The gravitational potential is $U(r) = -\frac{GM_{L}}{r},$ where $M_{L}$ is the lens mass.  Integration of equation (\ref{eq:Fermat-th}) yields  
\begin{widetext}
 \begin{equation}
\label{eq:Fermat-lensplane}
  \mbox{\cal{L}} =   \left(\frac{1}{2} (1 -\frac{\chi^2 \mu^2}{2p_0^2}) (\frac{1}{D_{OL}} + \frac{1}{D_{LS}})(\zeta-\eta)^2-4GM_{L} \ln{\zeta} +L_0\right) ,
 \end{equation} 
 \end{widetext}
where $L_0$ is a constant. 

The travel time is given by the same equation with just a sign change
\begin{widetext}
 \begin{equation}
\label{eq:traveltime-lensplane}
  T =   \left(\frac{1}{2} (1 +\frac{\chi^2 \mu^2}{2p_0^2}) (\frac{1}{D_{OL}} + \frac{1}{D_{LS}})(\zeta-\eta)^2-4GM_{L} \ln{\zeta} +T_0(\mu^2)\right) ,
 \end{equation} 
 \end{widetext}
 where 
 \begin{equation}
 T_{0}(\mu^2) =  (1 +\frac{\chi^2 \mu^2}{2p_0^2}) D_{OS} + 2GM_{L}\ln{(D_{OL}D_{LS})}
 \end{equation}





 The "lens equation" gives the position of images. It is obtained by differentiating equation (\ref{eq:Fermat-lensplane}) with respect to $\zeta$. One gets the familiar equation
 \begin{equation}
\label{eq:lensBKogan}
  (\zeta-\eta)-\frac{r_{E}^2}{\zeta} = 0
 \end{equation} 
 where the Einstein radius $r_{E}$ is defined by 
  \begin{equation}
\label{eq:EinsteinBKogan}
  r_{E}^{2}(\mu) = \frac{4GM_L D_{OL}D_{LS}}{c^2 D_{OS} (1 - \frac{\chi^2 \mu^2}{2p_0^2})} \simeq \frac{4GM_L (1+ \frac{\chi^2 \mu^2}{2p_0^2})D_{OL}D_{LS}}{c^2 D_{OS}}.
 \end{equation}

Comparing with the geometric derivation of the lens equation (\citet{1992grle.book.....S}), it is clear that the light deflection angle is
\begin{equation} 
\psi(\zeta) = \frac{4GM_L (1+ \frac{\chi^2 \mu^2}{2p_0^2})}{c^2\zeta}
\end{equation}

The lens equation (\ref{eq:lensBKogan}) has solutions $\zeta_{\pm}$ defined by
\begin{equation}
\zeta_{\pm} = \frac{1}{2}(\eta\pm \sqrt{\eta^2+4r_E^2})
\end{equation}
Equation (\ref{eq:lensBKogan}) and (\ref{eq:EinsteinBKogan}) show that mixed photons are lensed like massive particles of mass $\chi \mu.$  However, as shown below (equation (\ref{eq:spread1})), there are  differences  between mixed photon lensing and ordinary massive photon lensing.

In the experiment of gravitational deflection of light by the Sun, one of the light rays, say $\zeta_{-}$ is occulted by the Sun. Due to the coupling of the photon to the HP, the apparent position of the source on the sky moves by 
\begin{equation}
{\Delta \theta}_{\mathrm coh} = \frac{\Delta \zeta_{+}}{D_{OL}} =  \frac{\chi^2 \mu^2 {r_{E}}^2(0)}{2p_0^2 \sqrt{\eta^2+4{r_{E}}^2(0)} D_{OL}} 
\label{eq:spatialshift}
\end{equation}
relative to the position expected for massless photons. $r_{E}(0)$ is a short-hand notation for the Einstein radius evaluated at $\mu = 0.$ 

The signal of a mixed photon with energy $p_0$ is delayed relative to the massless photon arrival by
\begin{equation}
\Delta T_ {\mathrm coh}= \frac{\chi^2 \mu^2}{2p_0^2} D_{OS}(1+\frac{\zeta_{-}^2}{2D_{OL}D_{LS}})
\label{eq:temporalshift}
\end{equation}

Oscillations are only possible if the photon/HP beam has a non-vanishing momentum width, at least of order $p_{0}-\sqrt{p_{0}^2-\mu^2} \simeq \frac{\mu^2}{2p_{0}}.$ The $\zeta_{+}$ image is smeared on the sky with a width of at least 
\begin{equation}
\label{eq:spread1}
\frac{\delta(\Delta \zeta_{+})}{\Delta \zeta_{+}} \sim \frac{\mu^2}{p_{0}^2}.
\end{equation} 
The travel time $ \Delta T_{\mathrm coh}$ from equation (\ref{eq:temporalshift}) also obtains a relative dispersion of at least $\frac{\mu^2}{p_{0}^2}.$ By contrast, a lensed massive particle observed by an ideal 
telescope would be seen as a point-like image, with an arbitrarily small arrival time spread.

\section{Coherent and incoherent lensing}\label{sec:coherence}

If the photon and the HP were not mixed, they would move at different speeds on different geodesics. They would give 2 sets of images with different travel times (as noted by \citet{1981PhRvD..24..110K}).  
 The angular distance between say the $\zeta_{+}$ images would be given by 
\begin{equation}
\Delta \theta_{\mathrm incoh} = \frac{\Delta \zeta_{+}}{D_{OL}}=  \frac{\mu^2 {r_{E}}^2(0)}{2p_0^2 \sqrt{\eta^2+4{r_{E}}^2(0)}D_{OL}}. 
\label{eq:spatialshift2}
\end{equation}
for photons of energy $p_{0}.$ 
The time of arrival of the wave packets would be separated by 
 \begin{equation}
\Delta T_{\mathrm incoh} = \frac{\mu^2}{2p_0^2} D_{OS}(1+\frac{\zeta_{-}^2}{2D_{OL}D_{LS}}) \simeq \frac{\mu^2}{2p_0^2} D_{OS}
\label{eq:temporalshift2}
\end{equation} 
For the oscillation to occur, the photon and HP wave packets have to overlap temporally and spatially. For neutrino beams, this implies the existence of a coherence length \citep{1976PhLA...63..201N}, as discussed in numerous papers \citep{1981PhRvD..24..110K,1998PhRvD..58a7301G}.  From equations (\ref{eq:temporalshift2}) and (\ref{eq:spatialshift2}), the time overlap of the photon-HP wave packets is possible if 
\begin{equation}
\Delta T_{\mathrm incoh}  \simeq \frac{\mu^2}{2p_0^2} D_{OS} < \sigma_{x}
\label{eq:cohcond1}
\end{equation}
 where $\sigma_x$ is the largest among the temporal and spatial coherence width of the production and detection process \citep{1998PhRvD..58a7301G}.  If the condition (\ref{eq:cohcond1}) is satisfied, then the condition for space overlap of the wave packets
  \begin{equation}
D_{OL} \Delta \theta \sim r_{E} \frac{\mu^2}{2p_0^2} < \sigma_{x}
\label{eq:cohcond1}
\end{equation}
 is also fulfilled, since $r_{E} \ll D_{OS}.$

We now specialize to the observations of the gravitational deflection of radio waves by the sun. \citet{2009ApJ...699.1395F}  observed the relative position of 3C 279  (redshift z=0.536, angular size distance $D_{OS}=1.27 \mathrm{Gpc}$) and 3 fainter quasars with the VLBA. The central frequency of observations was  43 GHz and the passband width was 16 MHz. A limit of $\mu < 1.7\ 10^{-5} \mathrm{eV}$ has been set on the photon mass based on these observations
(\citet{2010IJMPD..19.2393A}). 
One would naively expect that this limit (obtained by exploiting equation (\ref{eq:spatialshift})) would translate directly into a limit on the effective mass $\mu \chi$ of the photon in the HP model. 
However, this is true only when the distance to the source $D_{OS}$ is larger than the coherence length $\Lambda_{\mathrm{coh}}.$
The coherence length has been evaluated for radio sources, especially Active Galactic Nuclei  by  \citet{2013PhRvD..87f5004L} in the context of photon-hidden photon mixing, 
\begin{equation}
\Lambda_{\mathrm{coh}} = 4\sqrt{2} \sigma_x (\frac{\omega}{\mu} )^2 =  4\sqrt{2} \sigma_x  ((\frac{\omega}{43 \mathrm{GHz}}) (\frac{1.8\ 10^{-4} \mathrm{eV}}{\mu}))^2.
\label{eq:cohlength} 
\end{equation}

Since synchrotron radiation from a single electron in the quasar source has a broadband spectrum, the temporal coherence width is dominated by the detection process, giving a lower limit for $\sigma_x $ of 19 meters. 
Inserting this value into equation (\ref{eq:cohlength}) shows that the condition $\Lambda_{\mathrm{coh}} > D_{OS}$ can be satisfied only for $\mu < 10^{-16} \mathrm{eV}.$
This tiny value is smaller than the photon mass in the interstellar medium ($10^{-13} \mathrm{eV}$) or the intergalactic medium plasmas  ($10^{-14} \mathrm{eV}$) \cite{2013PhRvD..87f5004L}. Plasma effects would hide totally any additional photon mass-induced deflection for HP masses smaller than $10^{-13} \mathrm{eV}$.  
In the $10^{-12} \mathrm{eV}<\mu < 10^{-5} \mathrm{eV}$ mass range,   the HP mass is larger than the plasma-induced photon mass and smaller than $\omega.$ Since the photon-HP beam is incoherent, a HP signal would be weak (weaker than the photon signal by $\chi^2$) and delayed relative to the photon signal by $\sim 10\ (\frac{\mu}{10^{-12}})^2\mathrm{s}.$ 3C279 is variable with a timescale of $\sim 1\ \mathrm{day}$ at radio frequencies \citep{2017MNRAS.464..418R}. A 10-day observation similar to  that of \citet{2009ApJ...699.1395F}
could in principle put an upper limit on the coupling $\chi$ for hidden photon masses in the $10^{-10} \mathrm{eV}<\mu < 10^{-9} \mathrm{eV}$ range.  


\section{Conclusion}

The hidden-photon model is a convenient toy model to study the lensing of oscillating particles. The photon-HP oscillations induce a refractive index for the photon. The deflection angle and time delay are then obtained with Synge's formalism.
The photon-HP beam  is lensed almost like a single massive particle of mass $\chi \mu$ as long as the photon and HP wave packets overlap. The only differences with massive lensing are a tiny angular spread of the images and spread in arrival time.
When the distance to the photon source is larger that the coherence length of the wave packets, the photon and the HP are lensed separately and give 2 sets of spatially and temporally separated images. 
The results obtained in this paper should extend qualitatively to more complex problems such as neutrino lensing.    
  
\begin{acknowledgements}
I would like to thank Jim Rich for stimulating and helpful comments. 
\end{acknowledgements}

\bibliography{hiddenphotonlensing}

\end{document}